\documentclass[a4paper,12pt]{article}
\usepackage{graphicx,amssymb,bm}

\newcommand{\vs}[1]{\vspace{#1 mm}}

\begin{document}
\topmargin -10mm
\oddsidemargin 0mm

\renewcommand{\thefootnote}{\fnsymbol{footnote}}

\begin{titlepage}
\setcounter{page}{0}

\vs{10}
\begin{center}
{\Large \bf United Approach to the Universe Model and the Local
Gravitation Problem}
\vs{15}

Guang-Wen Ma\footnote{e-mail address: guangwenma@sohu.com}$^{a}$
and
Zong-Kuan Guo\footnote{e-mail address: guozk@itp.ac.cn}$^{b}$ \\
\vs{6}
{\footnotesize{\it 
 $^a$ Department of Physics, Zhengzhou University, 
      Zhengzhou, Henan 450052, PR China \\
 $^b$ Institute of Theoretical Physics, Chinese Academy of Sciences,
      P.O.Box 2735, Beijing 100080, China \\}}
\end{center}
\vs{15}

\begin{abstract}
A united approach of the large-scale structure of a
closed universe and the local spherically symmetric gravitational 
field is given by supposing an appropriate boundary condition. The 
general feature of the model obtained are the following. The universe 
is approximately homogeneous and isotropic on the average on large 
scale and is expanding at present, as described by the standard model;
while locally, the small exterior region of a star started long ago to
contract, as expected by the gravitational collapse theory.
\end{abstract}

\end{titlepage}
\newpage
\renewcommand{\thefootnote}{\arabic{footnote}}
\setcounter{footnote}{0}

\section{Introduction}

In the traditional approaches of the universe model and the local
gravitation problem, the two subjects are alone treated
respectively. In various universe models people only consider the
average structure on large scale; the distribution of matter is
imagined as homogenized or smoothed. A galaxy is nothing but a
perfect fluid element. On the other hand, when people investigate
local gravitation problem such as the spherically symmetrical
gravitational field, they often regard the region as
asymptotically flat spacetime, as if there is no remain matter in the
universe. True, such a treatment has its reasons: (i) the model is
simple, and it does not impair the knowledge of the large-scale
structure of the universe to ignore its minor details. (ii) For a
local gravitation problem, the Birkhoff's theorem~\cite{GB} seems to
be the basis for leaving the remaining matter of the universe out of
consideration. However, no matter how reasonable it is apparently,
the asymptotically flat behavior of the solution of local
gravitational field does not be in tune with a closed universe
model, while the real universe most likely is closed.

The cosmological redshift shows that the distances between
galaxies are
increasing, that is, the universe is expanding on the 
scale of the
space between galaxies at present. On the other hand, 
the existing
main form of the cosmic matter is the celestial body, some of
which are very compact. This shows that in some very small areas
of the expanding universe the space has begun to contract long
ago. To our knowledge, there is yet no model to treat the two
problems concurrently at present. In the present article we try to
give, by means of the investigation a spherically symmetric
gravitational field in an expanding closed universe, a united
approach to the universe model and a local gravitational problem.

In order to make it practical, we shall adopt a compromise
proposal, which can be called ``homogeneous by areas model" of
matter distribution, that is, the universe can be divided into a
few spherically symmetric parts with the same symmetric center, and
in each part the matter density is a constant.

\section{The Gravitational Field Equations and Their General Solutions}

Imagine in a closed Friedmann universe a local dense-matter area
appears owing to a certain kind of mechanism. For simplicity,we
assume that
its density is a constant. Of course,
around the dense-matter area a spherically 
symmetric rare-matter or
vacuum area may appear. Such a matter
distribution model is our
start point. In this case, the 3-dimension
spacelike
hypersurface, $t=$ constant, is a rotational hypersurface
embedded in a 4-dimension flat spacetime. In comoving coordinates 
the line element can be written
\begin{equation}
\label{eq1}
ds^2=dt^2-U(\psi ,t)d\psi ^2-V(\psi ,t)\sin ^2\psi (d\theta ^2+
     \sin ^2\theta d\phi ^2)
\end{equation}
where $0\le\psi\le\pi$, $0\le\theta\le\pi$, $0\le\phi\le 2\pi$.
The relation between
4-dimension spherical coordinates $R$, $\psi$, $\theta$ and $\phi$ 
and the Decare
coordinates $x$, $y$, $z$ and $w$ are
\begin{eqnarray}
x & =& R\sin\psi\sin\theta\cos\phi \nonumber \\
y & =& R\sin\psi\sin\theta\sin\phi \nonumber \\
z & =& R\sin\psi\cos\theta \nonumber \\
w & =& R\cos\psi
\end{eqnarray}
One often introduces
\begin{equation}
\label{eq3}
r=\sin\psi
\end{equation}
and rewrites (\ref{eq1}) as the form
\begin{equation}
ds^2=dt^2-A(r,t)dr^2-B(r,t)(d\theta ^2+\sin ^2\theta d\phi ^2)
\end{equation}
which is a well-known form~\cite{SW}.

However, it is necessary to emphasize
that $r$ is not a good coordinate owing to non-monodromy of the
transformation (\ref{eq3}). The same value of $r$ corresponds to two
values of $\psi$, i.e., two positions in the universe. This means
that $A(r,t)$ and $B(r,t)$ should have two set physical solutions in
principle, which corresponds to the two areas $0\le\psi\le\pi/2$ and 
$\pi/2\le\psi\le\pi$,
respectively.

Consider the zero pressure perfect-fluid model
\begin{equation}
T^{\mu \nu}=\rho U^\mu U^\nu
\end{equation}
where $U^r=U^\theta =U^\phi =0$, $U^t=1$. 
The Einstein's equations are~\cite{GB}
\begin{eqnarray}
\label{eq6}
\frac{1}{A}\Big(\frac{B''}{B}-\frac{B'^2}{2B^2}-\frac{A'B'}{2AB}\Big)
-\frac{\ddot{A}}{2A}+\frac{\dot{A}^2}{4A^2}-\frac{\dot{A}\dot{B}}{2AB}
&=& -4\pi G\rho \nonumber \\
-\frac{1}{B}+\frac{1}{A}\Big(\frac{B''}{2B}-\frac{A'B'}{4AB}\Big)-
\frac{\ddot{B}}{2B}-\frac{\dot{A}\ddot{B}}{4AB}
&=& -4\pi G\rho \nonumber \\
\frac{\ddot{A}}{2A}+\frac{\ddot{B}}{B}-\frac{\dot{A}^2}{4A^2}-
\frac{\dot{B}^2}{2B^2} &=& -4\pi G\rho \nonumber \\
\frac{\dot{B}'}{B}-\frac{\dot{B}B'}{2B^2}-\frac{\dot{A}B'}{2AB} &=& 0 
\end{eqnarray}
From the last equation of (\ref{eq6}), it is easy to find that
\begin{equation}
\label{eq7}
A=\frac{B'^2}{4(1-Kr^2)B}
\end{equation}
here $K=K(r)$ is an arbitrary function of $r$. Using the firt three 
equations of  (\ref{eq6}) in (\ref{eq7}) gives that
\begin{equation}
\label{eq8}
4\frac{\ddot{B}\dot{B}^2}{B^3}+\frac{4Kr^2}{B}=0
\end{equation}
Set
\begin{equation}
\label{eq9}
B=S^2(r,t)r^2
\end{equation}
equation (\ref{eq8}) becomes
\begin{displaymath}
2\frac{\ddot{S}}{S}+\frac{\dot{S}^2}{S^2}+\frac{K}{S^2}=0
\end{displaymath}
from which we find
\begin{equation}
\label{eq10}
\dot{S}^2=\frac{2KF}{S}-K
\end{equation}
here $F=F(r)$ is an arbitrary function of $r$. The solution of 
equation (\ref{eq10}) is cycloid-like
\begin{eqnarray}
\label{eq11}
S &=& F(1-\cos\alpha) \nonumber \\
C+t &=& F(\alpha -\sin\alpha)/\sqrt{K}
\end{eqnarray}
where $C=C(r)$ is an arbitrary function. Substituting equations
(\ref{eq7}),
 (\ref{eq9}) and (\ref{eq11}) into (\ref{eq6}) gives the
expression of density
 
\begin{equation}
\label{eq12}
\rho=\frac{3}{4\pi G}\frac{(FKr^3)'}{S^3r^3}
\end{equation}

\section{Homogeneous by Areas Model of the Distribution of Matter}

Where we investigate the problem of spherically symmetric
gravitational collapse of a celestial body in a closed universe,
the "homogeneous by areas model" above is a suitable model. This
model finds expression in the form of the density, that is
\begin{equation}
\label{eq13}
\rho(r,t)=\left\{\begin{array}{llr}
          \rho_s(t) & 0\le\psi\le\psi_1 & (\textrm{area}\ I) \\
          0 & \psi_1\le\psi\le\psi_2<\pi/2 & (\textrm{area}\ II) \\
          \rho_u(t) & \psi_2<\psi\le\pi & (\textrm{area} III)           
          \end{array} \right.          
\end{equation}
where $\psi_1=\arcsin r_1$, $\psi_2=\arcsin r_2$. 
The joint conditions on the boundaries $\psi=\psi_1$ and $\psi=\psi_2$ 
is taken as that the metric  $g_{\mu \nu}$ are continuous on
these boundaries. Substituting the expression
(\ref{eq13}) into
equation (\ref{eq12}) and integrating gives 
\begin{equation}
\label{eq14}
\left\{ \begin{array}{lr}
          K_I=\frac{4\pi G}{3}\rho_sF_I^2(1-\cos\alpha_I)^3 
                 & (\textrm{area}\ I) \\
          K_{II}= \lambda / (F_{II}r^3) & (\textrm{area}\ II) \\
          K_{III}=\frac{4\pi G}{3}\rho_uF_{III}^2(1-
                 \cos\alpha_{III})^3 & (\textrm{area} III)
         \end{array} \right.
\end{equation}
where $\lambda$ is the integral constant for the area $II$; for the 
area $I$ and the area $III$, the integral constants have been taken
as zero so that both $K_I$ and $K_{III}$ are finite at $r=0$ 
(i.e., $\psi=0$ and $\psi=\pi$). The joint conditions are
\begin{eqnarray}
\label{eq15}
A_I(r_1,t)=A_{II}(r_1,t), \quad A_{II}(r_2,t)=A_{III}(r_2,t)\nonumber\\
B_I(r_1,t)=B_{II}(r_1,t), \quad B_{II}(r_2,t)=B_{III}(r_2,t)
\end{eqnarray}
The value of the parameter $\alpha$ in equation (\ref{eq11}) decides
that the scale factor $S(r,t)$ is expansive or contractive. As an
additional condition, we demand that
\begin{equation}
\label{eq16}
\alpha_I(r_1,t)=\alpha_{II}(r_1,t),
\quad \alpha_{II}(r_2,t)=\alpha_{III}(r_2,t)
\end{equation}
The expressions (\ref{eq15}) and (\ref{eq16}) are determining solution
conditions to $F(r)$, $K(r)$ and $C(r)$. From (\ref{eq15}) we can infer
that $S(r_1,t)$ is continuous on the boundaries and then from the first
of (\ref{eq11}) and (\ref{eq16}) we can know
\begin{equation}
\label{eq17}
F_I(r_1)=F_{II}(r_1), \quad F_{II}(r_2)=F_{III}(r_2)
\end{equation}
The second of (\ref{eq11}) can be written as
\begin{equation}
\label{eq18}
t-t_0=F_i\Big[(\alpha_i-\sin\alpha_i)-(\alpha_{i0}
-\sin\alpha_{i0})\Big]/\sqrt{K_i}
\end{equation}
here $i=I,II,III$ and $\alpha_{i0}=\alpha_i(r,t_0)$. Equations
(\ref{eq16}), (\ref{eq17}) and (\ref{eq18}) imply
\begin{equation}
\label{eq19}
K_I(r_1)=K_{II}(r_1), \quad K_{II}(r_2)=K_{III}(r_2)
\end{equation}
Substituting (\ref{eq16}), (\ref{eq17}) and (\ref{eq19}) into
(\ref{eq14}) gives
\begin{eqnarray*}
\lambda & = & \frac{4\pi G}{3}\rho_sF_I^3(r_1)r_1^3
              \Big(1-\cos\alpha_I(r_1,t)\Big)^3 \\
        & = & \frac{4\pi G}{3}\rho_uF_{III}^3(r_2)r_2^3\Big(1-
              \cos\alpha_{III}(r_2,t_1)\Big)^3 
\end{eqnarray*}
From the first and the last of (\ref{eq14}) it is also easy to see that
\begin{eqnarray}
\label{eq20}
\rho_s(1-\cos\alpha_I)^3 &=& \rho_{s0}(1-\cos\alpha_{I0})^3 \nonumber \\
\rho_u(1-\cos\alpha_{II})^3 &=& \rho_{u0}(1-\cos\alpha_{II0})^3
\end{eqnarray}

In general, these conditions can yet not determine $F_i$ and $K_i$ (or
$\alpha_{i0}$) completely because
there
 still exists a certain freedom of coordinate transformation in
comoving coordinates. In this sense, to select the function form
of $F_i$ and $\alpha_{i0}$ is just to select coordinates system.

Now we construct
a set of concrete solutions from the following considerations:
(i) In the case $\psi_2=\psi_1$ ($\rho_s=\rho_u$, see (\ref{eq20})), 
the universe should be
homogeneous and isotropic. 
(ii) In the case $\psi\gg\psi_1$, $\alpha_{II0}$ and $\alpha_{III0}$
should
be approximate constants and so are $F_{II}$ and $F_{III}$. 
(iii) In the case $\psi\ll\psi_2$ both $\alpha_{I0}$ and $\alpha_{II0}$ 
should be larger than $\pi$, so that the
scale factor in this area has been contracted, which reflects
gravitational collapse of a celestial body. The first condition
implies $\alpha_{I0}=\alpha_{II0}=\alpha_{III0}=$ constant, and 
$F_I=F_{II}=F_{III}=$ constant.
We try to choose the following forms of $\alpha_{i0}$ and $F_i$
\begin{equation}
\label{eq21}
\alpha_{i0}=\beta+\frac{r_2-r_1}{r_2+r_1}\delta
e^{-\psi/\sqrt{\psi_1\psi_2}}
\end{equation}
\begin{equation}
\label{eq22}
F_i= a-\frac{r_2-r_1}{r_2+r_1}be^{-\psi/\sqrt{\psi_1\psi_2}}
\end{equation}
where $\beta$, $\delta$, $a$ and $b$ are constants. Obviously, the 
last of (\ref{eq15}) is satisfied because of the conditions
(\ref{eq16}), (\ref{eq17}) and (\ref{eq19}). 
In order to
make the first of (\ref{eq15}) held,
we differentiate (\ref{eq18}) with respect to $r$ and
substitute into (\ref{eq14}). So doing, we obtain
\begin{eqnarray}
(1-\cos\alpha_I)\alpha_I' &=& \bigg\{(1-\cos\alpha_{I0})+
         \frac{3\sin\alpha_{I0}}{2(1-\cos\alpha_{I0})}\Big[(\alpha_I-
         \nonumber \\
       &&\sin\alpha_I)-(\alpha_{I0}-\sin\alpha_{I0})\Big]\bigg\}
          \alpha_{I0}'  \nonumber \\
(1-\cos\alpha_{II})\alpha_{II}' &=& (1-\cos\alpha_{II0})\alpha_{II0}'
         +\frac{3(F_{II}r)'}{2F_{II}r}\Big[(\alpha_{II}- \nonumber \\
       &&\sin\alpha_{II})-(\alpha_{II0}-\sin\alpha_{II0})\Big]
         \nonumber \\
(1-\cos\alpha_{III})\alpha_{III}' &=& \bigg\{(1-\cos\alpha_{III0})+
         \frac{3\sin\alpha_{III0}}{2(1-\cos\alpha_{III0})}
         \Big[(\alpha_{III}- \nonumber \\
       &&\sin\alpha_{III})-(\alpha_{III0}-
         \sin\alpha_{III0})\Big]\bigg\}\alpha_{III0}' 
\end{eqnarray}
The continuity of $\alpha$ on the boundaries gives
\begin{eqnarray}
\frac{(F_{II}r)'}{F_{II}r}\bigg\vert_{r_1} &=& -\frac{\sin\alpha_{I0}}
         {1-\cos\alpha_{I0}}\alpha_{I0}'\bigg\vert_{r_1} \nonumber \\
\frac{(F_{II}r)'}{F_{II}r}\bigg\vert_{r_2} &=& -\frac{\sin\alpha_{III0}}
         {1-\cos\alpha_{III0}}\alpha_{III0}'\bigg\vert_{r_2}
\end{eqnarray}
or equivalent forms
\begin{eqnarray}
\label{eq25}
\frac{b\omega_1}{(a-b\omega_1)\sqrt{\psi_1 \psi_2}}\frac{1}{\sqrt{1-
          r_1^2}}+\frac{1}{r_1} =  
          \frac{\delta\omega_1\sin(\beta+\delta\omega_1)}
          {\Big[1-\cos(\beta+\delta\omega_1)\Big]\sqrt{\psi_1 \psi_2}}
          \frac{1}{\sqrt{1-r_1^2}} \nonumber \\
\frac{b\omega_2}{(a-b\omega_2)\sqrt{\psi_1 \psi_2}}\frac{1}{\sqrt{1-
          r_2^2}}+\frac{1}{r_2} =  
          \frac{\delta\omega_2\sin(\beta+\delta\omega_2)}
          {\Big[1-\cos(\beta+\delta\omega_2)\Big]\sqrt{\psi_1 \psi_2}}
          \frac{1}{\sqrt{1-r_2^2}} 
\end{eqnarray}
where
\begin{eqnarray}
\omega_1 &=& \frac{r_2-r_1}{r_2+r_1}e^{-\sqrt{\psi_1}/\sqrt{\psi_2}}
             \nonumber \\
\omega_2 &=& \frac{r_2-r_1}{r_2+r_1}e^{-\sqrt{\psi_2}/\sqrt{\psi_1}}
             \nonumber
\end{eqnarray}
If we assign any two of the four constants, say $\beta$ and $a$
(initial conditions), the other two can be determined by (\ref{eq25}),
and a set of uniquely definite solutions is thus found.

\section{Disussion}

From (\ref{eq21}) and (\ref{eq22}) it is easy to see that if $r_2=r_1$
then $\alpha_{i0}=\beta$ and $F_i=a$, which exactly is the homogeneous
and isotropic Friedmann
universe model.
In general, where $\psi\gg\psi_1$ and $\psi\gg\psi_2$, we have
$\alpha_{III0}\approx\beta$ and $F_{III}\approx a$, which shows 
that the spacetime far away from the star tends to become homogeneous
and isotropic. In the place where $\psi\ll\psi_2$ We have
\begin{eqnarray}
\alpha_{I0}\approx\beta+\frac{r_2-r_1}{r_2+r_1}\delta \\
F_I\approx a-\frac{r_2-r_1}{r_2+r_1}b
\end{eqnarray}
The existing observational data show that $\beta\approx\pi/2$. If
$(r_2-r_1)/(r_2+r_1)\delta >\pi/2$,
then the celestial body is collapsing.

In the traditional treatment of the problem of the spherically
symmetric gravitational collapse, the exterior region of the
star is considered infinite and empty~\cite{SW,OS,MW}. The surface of
the star is
only boundary. A puzzle arose out of this treatment, which
we arrate as follows. Suppose the surface of a star with constant
density is described by the equation
\begin{equation}
\frac{x^2+y^2}{a^2}+\frac{z^2}{b^2}=1
\end{equation}
in a certain coordinate system. Under the transformations
\begin{equation}
x'=x,\quad y'=y,\quad z'=az/b
\end{equation}
The surface equation becomes
\begin{equation}
x'^2+y'^2+z'^2=a^2
\end{equation}
in the new coordinate system. Then which symmetry has the
gravitational field, axisymmetric or spherically symmetric? In our
new treatment there is no such a puzzle owing to the existence of
the second boundary.

Birkhoff's theorem states that the solutions
of spherically symmetric and vacuum gravitational field equations
must be Schwartzschild. Hence the metric of the area $II$ should be
able to become Schwartzschild form by means of a certain
transformation. Such a transformation has been found, which is
\begin{eqnarray}
\bar{r} &=& F_{II}r(1-\cos\alpha_{II}) \nonumber \\
\bar{t} &=& 2GM\ln\frac{\sqrt{F_{II}r-GM}\sin\alpha_{II}+\sqrt{GM}(1+
             \cos\alpha_{II})}{\sqrt{F_{II}r-GM}\sin\alpha_{II}
             -\sqrt{GM}(1+\cos\alpha_{II})} \nonumber \\
            && \sqrt{\frac{F_{II}r-GM}{GM}}(F_{II}r+2GM)\alpha_{II}-
             F_{II}r\sqrt{\frac{F_{II}r-GM}{GM}}\sin\alpha_{II}
             \nonumber \\
            && -\int\sqrt{\frac{GM}{F_{II}r-GM}}
              \bigg[\sqrt{\frac{F_{II}^3r^3}{GM}}
              (\alpha_{II0}-\sin\alpha_{II0})\bigg]'dr
\end{eqnarray}

\end{document}